\documentclass[10pt,letterpaper,twocolumn]{article} %% two column, final layout

\usepackage{ol2}
\usepackage[draft]{hyperref}
\usepackage{amsmath}
\usepackage{setspace}
\begin{document}

\twocolumn[ %% activate for two-column option

\title{Fiber-based ultra-stable frequency synchronization using client-side, 1f-2f active compensation method}

%% For REVTeX it is possible to automate superscript and e-mail callouts with the superscriptaddress option; see REVTeX4 documentation.

\author {B. Wang,$^{1,2}$ X. Zhu,$^{1,3}$ C. Gao,$^{1,2}$ Y. Bai,$^{1,3}$ J. Miao,$^{1,3}$ and L.J. Wang$^{1,2,3,4,*}$}

\address{
$^1$Joint Institute for Measurement Science, Tsinghua University, Beijing 100084, China
\\
$^2$Department of Precision Instruments, State Key Laboratory of Precision Measurement Technology and Instruments, Tsinghua University, Beijing 100084, China\\
$^3$Department of Physics, Tsinghua University, Beijing 100084, China\\$^4$National Institute of Metrology, Beijing 100013, China\\
$^*$Corresponding author: lwan@tsinghua.edu.cn
}

\begin{abstract}We demonstrate a frequency synchronization scheme with the phase noise compensation function placed at the client site. One transmitting module hence can be linked with multiple client sites. As a performance test, using two separate 50 km fiber spools, we recover the 100 MHz disseminated reference frequencies at two remote sites, separately. Relative frequency stabilities between two recovered frequency signals of $2.8\times10^{-14}/s$ and $2.5\times10^{-17}/day$ are obtained. This scalable scheme is suitable for the applications of frequency dissemination with a star-topology, such as SKA and DSN. \end{abstract}

 ] %% activate for two-column option

Benefiting from innovations in modern atomic clock, time and frequency have been the most accurate and stable physical quantities which can be measured and controlled\cite{Katori,chou, Hinkley,Jun}. With significant progress of precise time and frequency synchronization\cite{Fujieda, Marra, Predehl, Wang, France, PK, NIST, Miao}, besides metrology\cite{Fujieda2}, more and more areas can share these powerful tools and covert the measurement of target parameters to the measurement of time and frequency. For example, in the applications of radio telescope array \cite{Cliche,Shillue} and deep space navigation (DSN) network\cite{NASA,NASA2}, the spatial resolution is converted to the phase resolution of received signals. In these applications, the short term phase (frequency) synchronization (1s to 100 s) between different dishes is required. This requirement cannot be satisfied by traditional synchronization method via satellite link, normally, which can only realize long term ($>$1000 s) frequency synchronization\cite{Bauch, Levine}. The recently developed fiber-based frequency synchronization method makes it possible and can realize the ultra-high stable phase (frequency) synchronization at the integration time from 1 s to $10^6 s$\cite{BOWANG}. To increase its availability and accessibility, several fiber-based multiple-access frequency dissemination schemes have been proposed and demonstrated\cite{Grosche, Gaochao, Bai, PTB, France2, poland}. However, there are still many unsolved problems. In practical applications, to satisfy different topological synchronization requirements, different frequency synchronization schemes are required.

As a typical example of radio astronomy telescope arrays, the square kilometer array (SKA) telescope has a high requirement on the phase (frequency) synchronization between hundreds of dishes in the first phase of SKA (SKA1)\cite{SKA1}. This requirement will be extended to thousands of dishes in SKA2. The atomic clock ensemble is located at the center station, and all dishes will be connected to it via fiber links. The reference frequency of center station will be disseminated to all dishes under a star topology. In this case, the space requirement for the frequency dissemination system in the center station is a key problem. For all current fiber-based frequency dissemination schemes, the phase noise detection and compensation functions are normally located at the transmitting site. In the case of SKA1 where hundreds of dish telescopes are planned, corresponding number of compensation modules need to be placed in the same control station. This would generate extraordinary space requirement, and cause unnecessary difficulties for future expansion.

In this letter, we propose and demonstrate a new frequency synchronization method with the phase noise compensation module placed at the client site. One transmitting module can thus be linked with multiple client sites, and the expansion of future receiving sites will not disrupt the structure of the central transmitting station. As a performance test, using two 50 km fiber spools, we recover the 100 MHz disseminated reference frequency signals at two remote sites, separately. Relative frequency stabilities between the two recovered frequency signals of $2.8\times10^{-14}/s$ and $2.5\times10^{-17}/day$ are obtained. With this scalable scheme, the space requirement for frequency dissemination system at the center station can be dramatically reduced for the applications of SKA, DSN network, and any other cases requiring a star-topology multiple-access dissemination.

\begin{figure*}[htb]
\centerline{\includegraphics[width=16 cm]{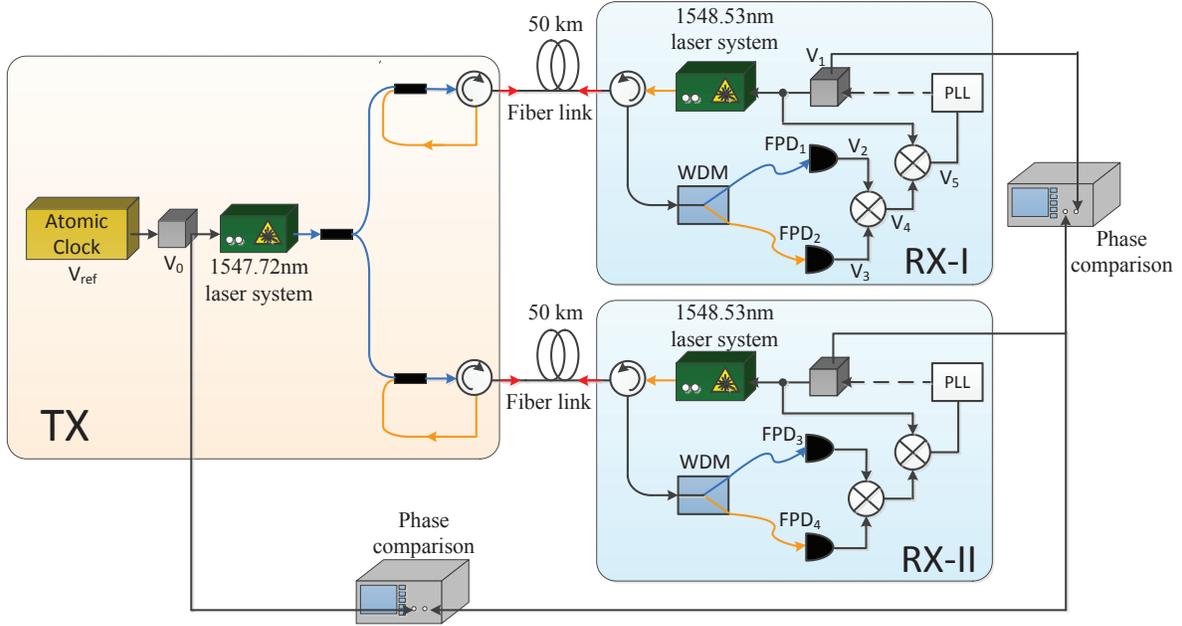}}
\caption{(Color online) Schematic diagram of the client side active compensated frequency dissemination system.}
\label{fig1}
\end{figure*}

Figure~\ref{fig1} shows the schematic diagram of the client side active compensated frequency dissemination system. One transmitting site (TX) is connected to two receiving sites (RX). As a performance test and for the convenience of phase difference measurement, the TX is linked with RX-I and RX-II via two spools of 50 km fiber, respectively. The function of TX is very simple - modulating and broadcasting. To increase the signal-to-noise ratio for compensation, the 100 MHz reference frequency ($V_{ref}$) from a Hydrogen maser (H-maser) is boosted to 2 GHz via a phase-locked dielectric resonant oscillator(PDRO), which can be expressed as $V_0=cos(\omega_0t+\phi_0)$ (without considering its amplitude). $V_0$ is used to modulate the amplitude of a 1547.72 nm diode laser. Taking the dissemination channel TX to RX-I as an example, after passing a ``2 to 1" fibre coupler and an optical circular, the modulated laser light is coupled into the 50 km fibre link. The structure of RX is more complex. A 1 GHz PDRO, which can be expressed as $V_1=cos(\omega_1t+\phi_1)$, is phase locked to a 100MHz oven controlled crystal oscillator (OCXO). The phase of OCXO can be controlled by an external voltage. $V_1$ is used to modulate the amplitude of a 1548.53 nm diode laser. With the help of two optical circulators at RX-I and TX site, the modulated 1548.53 nm laser light transfers in the same 50 km fiber spool via the route RX-I to TX to RX-I. After the round trip transfer, the 1548.53 nm laser light can be separated from the received one-way 1547.72 nm laser light by a wavelength division multiplexer (WDM). The modulated 1547.72 nm laser is detected by a fast photo-diode FPD1. The recovered 2 GHz frequency signal can be expressed as $V_2=cos(\omega_0t+\phi_0+\phi_p)$, where $\phi_p$ is the phase fluctuation induced during the 50 km fiber dissemination. The modulated 1548.53 nm laser carrier is detected by FPD2 and the recovered 1 GHz frequency signal can be expressed as $V_3=cos(\omega_1t+\phi_1+\phi_p')$, where $\phi_p'$ is the phase fluctuation induced by the 100 km fiber dissemination (round trip). We mix down the signals $V_2$ and $V_3$ to obtain $V_4=cos[(\omega_0-\omega_1)t+\phi_0+\phi_p-\phi_1-\phi_p']$. Then, mixing down the signals $V_1$ and $V_4$ gives an error signal $V_5=cos[(\omega_0-2\omega_1)t+\phi_0+\phi_p-2\phi_1-\phi_p']$. $V_5$ is used for the phase-locked loop (PLL) to control the phase of OCXO. As the relationship of $\omega_0/\omega_1=2$, via the same fiber link, the one-way accumulated phase fluctuation of the 2 GHz frequency signal is the same as the round-trip accumulated phase fluctuation of the 1 GHz frequency signal, $\phi_p=\phi_p'$. Consequently, the DC error signal can be expressed as $V_5=cos(\phi_0-2\phi_1)$. When the PLL is closed, the phase of $V_1$ at RX-I will be locked to the phase of $V_0$ at TX, and realizing the stable-frequency dissemination from TX to RX-I. The same structure is applied to the RX-II and all other sites. The method is sometimes referred to as the ``1f-2f" method due to the factor-2 relationship between the TX and RX frequencies.

\begin{figure}[htb]
\centerline{\includegraphics[width=8.5 cm]{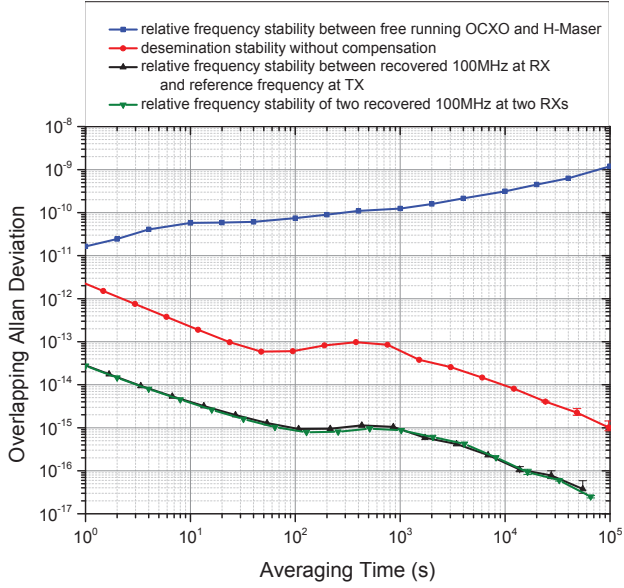}}
\caption{(Color online) Measured frequency stability of the dissemination system without compensation (red curve); relative frequency stability between free running OCXO and H-Maser(blue curve); relative frequency stability between recovered 100MHz at RX and reference frequency at TX (dark curve) when the PLL is closed; relative frequency stability of two recovered 100MHz at two RXs when two PLLs are closed (green curve).}
\label{fig2}
\end{figure}

To test the dissemination stability of the proposed client site active compensation method, we perform a series of measurements.The results are shown in Fig.~\ref{fig2}. The blue curve is the measured relative stability of 100 MHz frequency signals between the H-maser at TX and OCXO at RX-I when the PLL is open. The relative frequency stabilities of $1.6\times10^{-11}/s$ and $1.2\times10^{-9}/day$ depend on the OCXO we used. The red curve shows the relative frequency stability between $V_0$ and $V_2$ when the PLL is open. The relative frequency stabilities of $2.5\times10^{-12}/s$ and $1\times10^{-15}/day$ represent the 50 km fiber dissemination stability without compensation. The dark curve is the measured relative stability of 100 MHz frequencies between H-maser at TX and OCXO at RX-I when the PLL is closed. The relative frequency stabilities of $3\times10^{-14}/s$ and $3.8\times10^{-17}/day$ have been realized, which means the OCXO at RX-I has been phase locked to the H-maser at TX. We also measure the relative frequency stability between two OCXOs at RX-I and RX-II when PPLs of two dissemination links are all closed (green curve). The relative frequency stabilities of $2.8\times10^{-14}/s$ and $2.5\times10^{-17}/day$ have been realized, which means the OCXOs at two RXs are phase coherent to each other within the compensation bandwidth of the dissemination system. Based on these measurements, we can see the proposed client side active compensation scheme can satisfy the reference frequency dissemination requirement on a star-topology fiber network.

In the proposed scheme, the two laser carriers' wavelengthes are different. Consequently, the fiber dispersion may cause a phase delay difference between the two laser carriers, and this may limit the frequency dissemination stability. The phase time delay $\tau(\lambda, T)$ of the disseminated frequency signal is
 \begin{equation}
 \tau(\lambda, T)=\frac{n(\lambda, T)\cdot L}{c},
 \end{equation}
 where c is the light speed in vacuum, $n(\lambda,T)$ is the fiber refractive index at optical wavelength $\lambda$ and temperature T, L is the fiber length. In practical application, temperature fluctuation of the fiber link is the dominant factor which will degrade the frequency dissemination stability. We calculate the second-order partial derivative of phase time delay
 \begin{equation}
\begin{split}
\frac{\partial^2\tau(\lambda, T)}{\partial\lambda \partial T}&=\frac{1}{c}\cdot\frac{\partial[\frac{\partial n(\lambda, T)}{\partial\lambda}\cdot L+\frac{\partial L}{\partial\lambda}\cdot n(\lambda, T)]}{\partial T}\\&=\frac{\partial [D(\lambda, T)\cdot L]}{\partial T}\\&=L\cdot\frac{\partial D(\lambda,T)}{\partial T}+D(\lambda,T)\cdot\frac{\partial L}{\partial T}\\&=L\cdot[ \kappa+D(\lambda, T)\cdot \alpha],
\end{split}
\end{equation}
with the fiber dispersion
\begin{equation}
D(\lambda,T)=\frac{1}{c}\cdot\frac{dn(\lambda,T)}{d\lambda},
\end{equation}
the chromatic dispersion thermal coefficient
\begin{equation}
\kappa=\frac{dD(\lambda,T)}{dT},
\end{equation}
and the fiber length expansion coefficient
\begin{equation}
\alpha=\frac{1}{L}\cdot \frac{dL}{dt}.
\end{equation}
For the commercial G652 fiber, $\alpha=5.6\times 10^{-7}/^o C$\cite{M}, $D=17ps/nm\cdot km$ and $\kappa=-1.45\times10^{-3}ps/(km\cdot nm\cdot^o C)$\cite{Zhong} around 1550 nm. Obviously, the chromatic dispersion thermal coefficient is the dominant item on the phase time delay difference. For a 100 km fiber dissemination, the phase time delay difference will be $\delta \tau=-0.14ps/nm\cdot^oC$. Supposing a diurnal temperature fluctuation of $30^oC$, the phase time delay difference of the frequency signal carried by 1547.72 nm and 1548.53 nm (0.81 nm difference) will be 3.5 ps, which corresponds to the dissemination stability of $8.1\times10^{-17}$ at the integration time of half day. This frequency dissemination stability can satisfy almost all current practical applications. Using the dense wavelength division multiplexing (DWDM) technique, two laser carriers' wavelength can be as near as 0.4 nm, and this will reduce the chromatic dispersion impact on the dissemination stability.

\begin{figure}[htb]
\centerline{\includegraphics[width=8.5 cm]{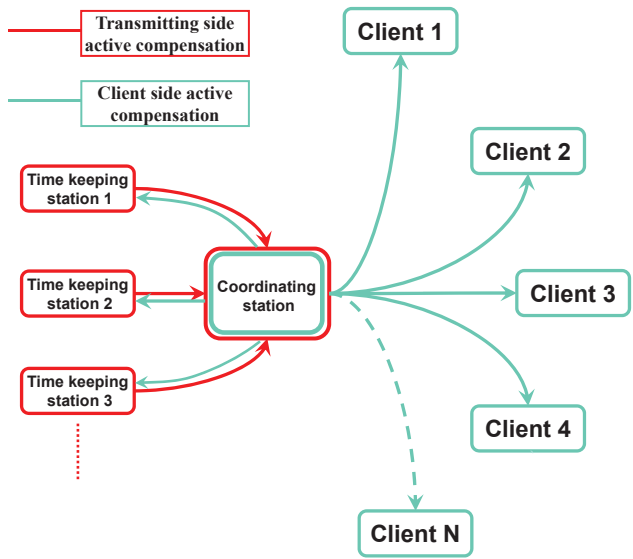}}
\caption{(Color online) Schematic diagram of a regional time and frequency synchronization network.}
\label{fig3}
\end{figure}
Currently, we are constructing the Beijing regional time and frequency synchronization network. According to functions, the synchronization network can be separated into three parts - time keeping stations, coordinating station and client sites, as shown in Fig.~\ref{fig3}. The time keeping points are the institutes which maintain the national or regional official times. Using the transmitting site active compensation scheme, a time keeping station can perform frequency comparison, and transmit its time-frequency signal to a central coordinating station. Here, the received time-keeping signals can be compared in real time to generate a coordinated time-frequency signal. The coordinated frequency can then be disseminated back to each time keeping stations and all user clients via the proposed client-side active compensation method. As one transmitting module can be linked with multiple clients, the expansion of future dissemination channels (expansion of the clients) will not have significant impact on the coordinating station's basic structure.

The authors acknowledge funding supports from the Major State Basic Research Development Program of China (No.2010CB922900) and the National Key Scientific Instrument and Equipment Development Projects (No.2013YQ09094303).

\pagebreak
\section*{Informational Fourth Page}
 {\bf Full versions of citations}

\end{document}